# Magnetic transport and chaotic orbits of charged particles

D. Dubbers

*Physikalisches Institut, Heidelberg University, Im Neuenheimer Feld 226, 69120 Heidelberg, Germany*

**ABSTRACT**

I study electron movement in electromagnetic fields beyond the adiabatic approximation, using so-called Størmer theory. Some of the electron orbits are regular or integrable, but their measure is zero. Other orbits, called quasiperiodic, are unstable, but only for infinite times. All other orbits are chaotic or "hyperchaotic", or simple scattering states. Examples for typical electron orbits are given for all these cases. An open question still is whether the spectral changes due to Størmer will translate into changes in the limits on neutrino masses and on errors in neutrino-electron correlation experiments.

# I. INTRODUCTION

When electrons (or any other charged particles) are moving in a constant magnetic field, then, in the simplest case, the electrons perform a helical cyclotron movement along a given field line of the so-called guiding field. I call this process magnetic transport of the electron from one location to another. When the guiding field is non-uniform, the process of magnetic transport may be quite complicated. In this case, often the adiabatic approximation is used, in which case the guiding field changes only slowly in the particle's reference frame, see standard textbooks on electrodynamics like Ref. [1]. Usually, a coefficient of adiabaticity is introduced, which for adiabatic transport is required to be much smaller than one, where the question remains open how much smaller this coefficient should be. The present paper discusses effects far beyond the adiabatic approximation, due to strongly non-adiabatic electron transport. Then I investigate whether the arrival of chaos theory during past few decades can improve our understanding of nonadiabatic effects.

The paper is organized as follows. Section II gives an overview to the problem of electron orbits in a static magnetic dipole field of any shape, mainly based on what is known as "the Størmer problem". Størmer's theory treats the electron as an imaginary particle moving in a space-depending potential involving a fixed saddle point, where the particle is lost to infinity as soon as its energy surpasses the potential of the saddle point. Furthermore, from Størmer theory I find some unexpected features of electron orbits in the effective potential that can only be understood in the context of chaos theory, not yet available during the lifetime of Størmer. Section III discusses chaos theory and its impact on Størmer.theory

## II. THE STØRMER PROBLEM

Deviations from the adiabatic approximation may strongly affect the orbits of charged particles, up to their uncontrolled losses. The present section discusses efforts to understand such strongly irregular electron orbits and losses, prior to the advent of chaos theory. This is done in some detail, because key papers on the subject are not easily accessible: often written in languages other than English, published in journals not easily accessible to physicists, using various notations, often with parametrization borrowed from geophysics, and sometimes having an insufficient list of references. This makes the access to the field difficult, and this may be the reason why Størmer's work is not widely known to physicists.

Historically, Størmer's main objective was to understand the generation of polar lights in the earth's dipolar magnetic field, assumed to be axially symmetric and constant in time. Størmer was a Norwegian mathematician well-known at the time for his work on number and gauge theories. Størmer worked on the polar light problem for more than 30 years, with about ten publications on the subject, almost all in German or French, which culminated in 1930, Ref. [2], in which first chaotic orbits are presented. This work was done well before the advent of chaos theory and of computers, and was only possible because Størmer was granted a battalion of unnamed mathematicians to do the calculations.

We start with Hamilton's theory for a given particle, subject to external fields. For a given Hamiltonian $\mathcal{H}_R$, the equations of motion are the Hamilton equations

$$\dot{q}_i = \frac{\partial \mathcal{H}_R}{\partial p_i}, \quad \dot{p}_i = -\frac{\partial \mathcal{H}_R}{\partial q_i}, \tag{1}$$

from which the particle's positions $q_i(t)$ and conjugate momenta $p_i(t)$ are derived. In Eq. (1), only the derivatives of $\mathcal{H}_R$ with respect to the phase space variables enter, and not the Hamiltonian itself, therefore one has a certain freedom in the choice of the Hamiltonian.

In our case of an electron propagating in a static (time independent) and rotationally symmetric magnetic field, a relativistic canonical Hamiltonian in three-dimensional cylindrical coordinates $\{z, \rho, \varphi\}$ is [3]

$$\mathcal{H}_R = \{m^2 c^4 + c^2 [p_z^2 + p_\rho^2 + (p_\varphi/\rho - eA_\varphi)^2]\}^{1/2}, \tag{2}$$

with electron charge $e$, mass $m = \gamma m_0$, Lorentz factor $\gamma = (1 - v^2/c^2)^{-1/2}$, and with speed $v$ given in terms of the speed of light $c$.

The electron's momentum components are $p_z = m\dot{z}$, and $p_\rho = m\dot{\rho}$. The field's angular momentum is a conserved quantity $p_\varphi = const$, due to the rotational symmetry of the problem. Therein, the $\varphi$-component of the vector potential for the dipole

$$A_\varphi = M_r \frac{\rho}{r^3}, \tag{3}$$

with reduced dipole moment $M_r = \frac{\mu_0}{4\pi} M$, vacuum magnetic permeability $\mu_0$, dipole moment $M$, and $r = (z^2 + \rho^2)^{1/2}$. In a static, that is, time independent magnetic vector potential $\mathbf{A}$ (with field $\mathbf{B} = \nabla \times \mathbf{A}$) this Hamiltonian is a constant of the motion

$$\mathcal{H}_R = \gamma m c^2 = const. \tag{4}$$

In terms of Hamiltonian theory, the Størmer problem can be rewritten as

$$\mathcal{H} = \frac{1}{2m}[p_z^2 + p_\rho^2 + \left(\frac{p_\varphi}{\rho} - eA_\varphi\right)^2], \tag{5}$$

$\mathcal{H}$ has the same partial derivatives with respect to the phase space variables $\{z, \rho, p_z, p_\rho\}$ as $\mathcal{H}_R$, and is also a constant of the motion,

$$\mathcal{H} = \frac{1}{2} m v^2 = const. \tag{6}$$

Most importantly, $\mathcal{H}$ finds a new interpretation as a classical mechanics problem, describing an imaginary particle moving in a two-dimensional effective potential $V$.

This effective potential, when written in dimensionless variables, with time measured in units of $t_0 = M_r/p_\varphi$, and length measured in units of $x_0 = mM_r^2/p_\varphi^3$, reads

$$V = \frac{1}{2}(\frac{1}{\rho} + \frac{\rho}{r^3})^2. \tag{7}$$

$V$ can be extended to comprise a static electric component $- e\, E\, z$.

Insertion of the partial derivatives of $\mathcal{H}$ into Eq. (1) then gives a system of differential equations for the phase space elements in the *meridian* plane (passing through the north and south poles) $\{z, \rho, p_z, p_\rho\}$, namely the Hamilton equations

$$\dot{z} = p_z, \quad \dot{\rho} = p_\rho,$$

$$\dot{p}_z = -\frac{3z\rho}{r^5}\left[\frac{1}{\rho} - \frac{\rho}{r^3}\right], \tag{8}$$

$$\dot{p}_\rho = \left[\frac{1}{\rho^2} + \frac{1-3\rho^2}{r^3}\right]\left[\frac{1}{\rho} - \frac{\rho}{r^3}\right].$$

For initial conditions $\{z_0, \rho_0, p_{z0}, p_{\rho 0}\}$, we find the solutions $\{z(t), \rho(t), p_z(t), p_\rho(t)\}$. (Motion in the *equatorial* plane is separately described by particle motion in a potential $V' = 1/(2\rho^2)(1 - 1/\rho)^2$, see Ref. [4], but I do not make use of this.). In cylindrical coordinates $\{r, \lambda\}$, with latitude $\lambda$ (the usual polar angle $\pi/2 - \lambda$ is called the colatitude), a dipole's field line in the meridian plane is given by

$$r = r_{0\,max} \cos^2 \lambda, \tag{9}$$

or, in $\{z, \rho\}$ coordinates,

$$z = \pm \rho^{2/3} (r_{0\,max}^{2/3} - \rho^{2/3})^{1/2}, \tag{10}$$

Fig. 1. Shows a field line, Eq.(9), from a dipole $M$ and an electron of momentum $p$ emitted along the field line

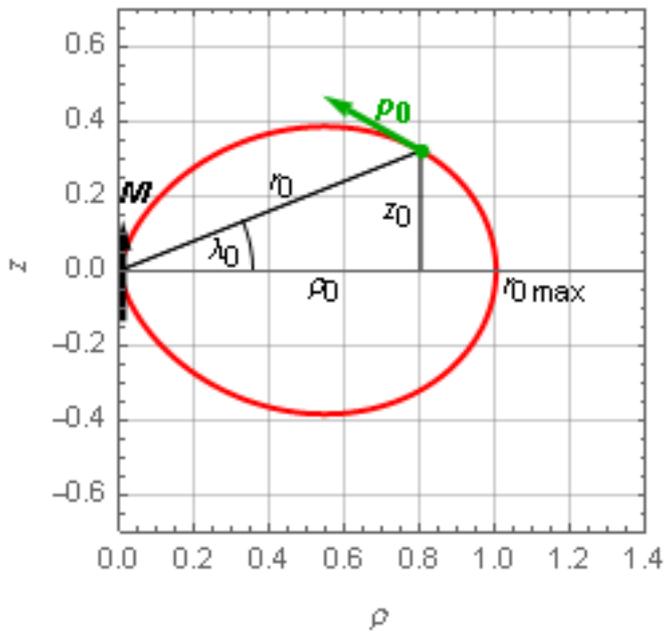

Fig. 1. In red, a field line in the meridional plane of a point dipole $M$, positioned at the origin and pointing along positive *z*, see. Eq. (10). Shown is he momentum $p$ of an electron (green arrow), emitted at position $\{\rho, z\}$ (green dot), along the field line.

For initial conditions $z_0 = 0$ and $\rho_0 = r_{0max} = 1$, the charged particle starts at what Størmer calls the potential's 'Thalweg' where $V = 0$, given by the line $\rho^2 = r^3$, or, equivalently, $z = \pm(1 - \rho^{2/3})^{1/2}$.

Fig. 2 shows Størmer's effective potential, Eq. (7), as seen by the electron.

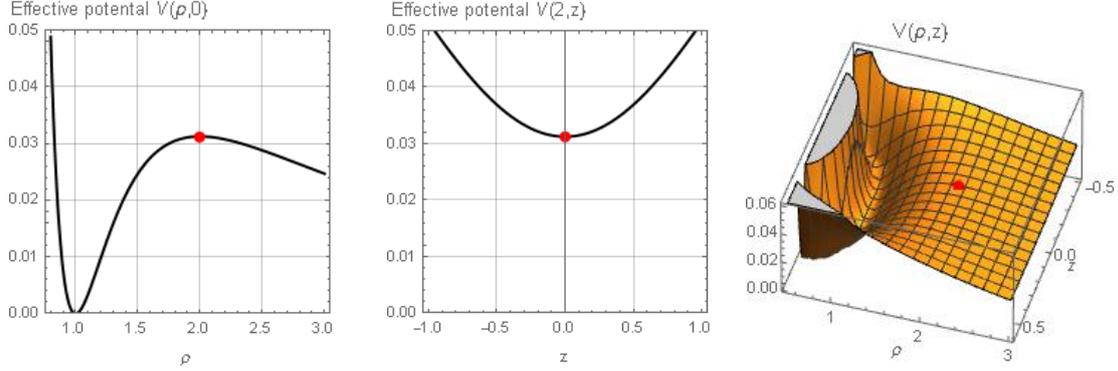

Fig. 2. The effective potential $V(\rho, z)$ of the Størmer problem has a saddle point at $z = 0$, $\rho = 2$ (red dot) where $V_{saddle} = 1/32$. Shown are the cuts $V(\rho, 0)$ at $z = 0$ (left panel) and $V(2, z)$ at $\rho = 2$ (middle panel), and the 3D presentation of $V(\rho, z)$.

In this effective potential, most electrons will perform irregular (chaotic) motions. But up to year 2000, publications on the Størmer problem were mostly limited to finding regular (integrable) orbits of the particles [5], [6], in the ocean of irregular orbits, simply because chaotic orbits were regarded as intractable. For a more recent experimental study on positron orbits in a magnetic field setup, see Ref. [7].

The existence of a saddle in the potential means that the trajectory of a particle launched within the valley (to the left of the saddle) with insufficient energy is not integrable, whereas the trajectory of a particle that surpasses the saddle point is integrable and, in the long run, reaches a straight line.

Fig. 3 first shows the development of the phase space variables $\rho(t)$, $z(t)$ of our (artificial) charged particles. From this the particle's orbits $z(\rho)$ are derived using a parametric plot. The third row of panels shows the orbits $z(\rho)$, which are then superimposed onto the potential $V(\rho,z)$ in the fourth row of panels. The panels in the vertical columns are, from left to right, for particles of low momentum $p_{z0} = 0.05$, (called "Quasiperiodic"), for particles of intermediate momentum $p_{z0} = 0.20$, (called "Hyperchaotic"), and for particles of higher momentum $p_{z0} = 030$, (called "Scattering"). Notations were chosen to be compatible with the next section on chaos theory.

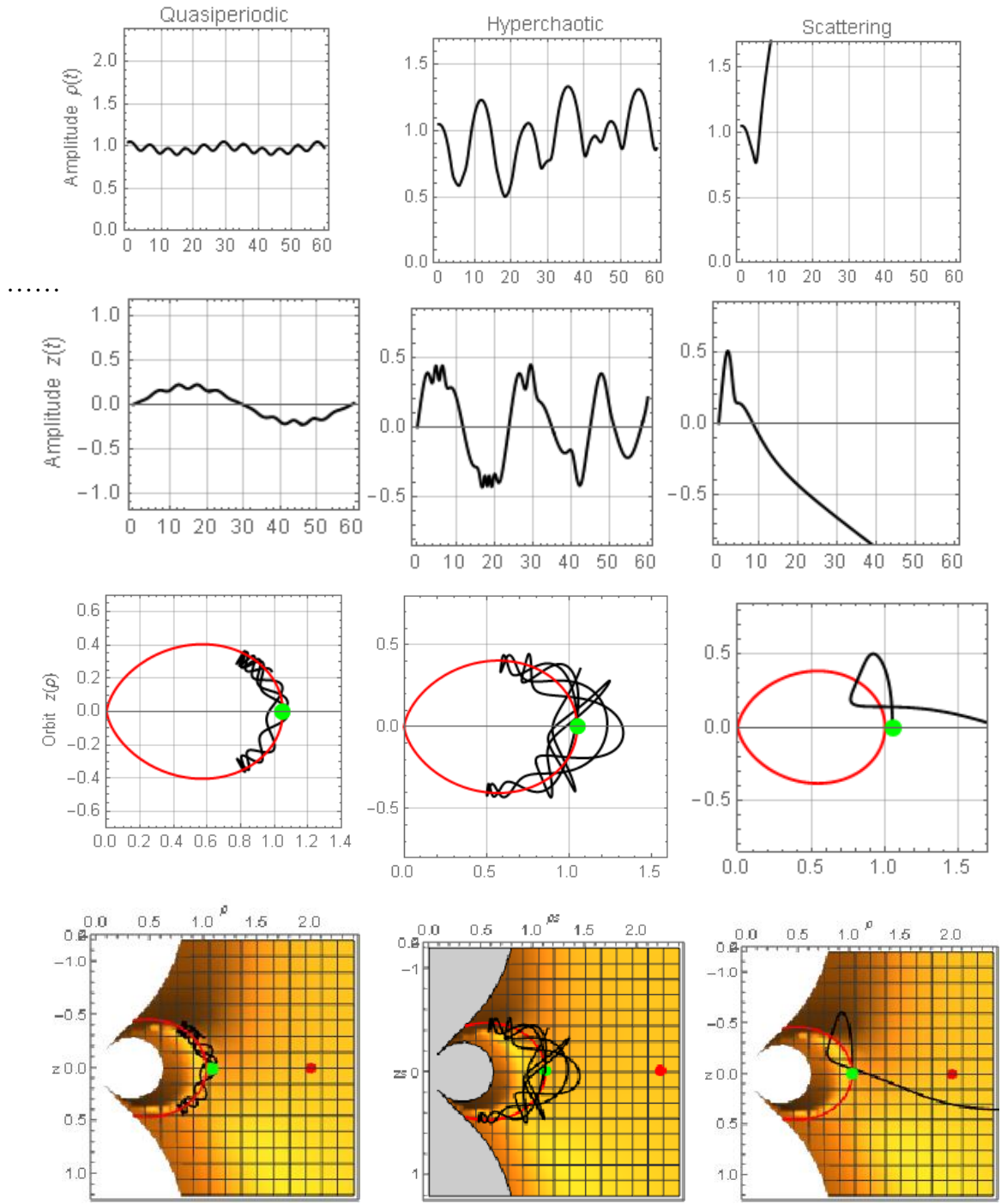

Fig. 3. Evolution of phase space variables of the charged particles, from initial conditions chosen to $\{z_0, \rho_0, p_{z0}, p_{\rho 0}\} = \{0, 1.05, p_{z0}, 0\}$. These particles are emitted under angle zero with respect to the red field line. The panels in the horizontal rows show first the amplitudes of positions $\rho(t)$ and $z(t)$ of our particles. The time parameter then is eliminated by using a parametric plot giving the electron orbit $z(\rho)$, see the third row of panels. This orbit then is superimposed on the effective potential from the righthand side of Fig. 2, shown in the fourth row of panels. Particles start at $t = 0$ on the green dots, while the red dot indicate the position $z = 0, \rho = 2$ of the saddle-point of the effective potential.

Next we look at a particle starting at a high field value near the origin and then going downhill, shown in Fig. 4.

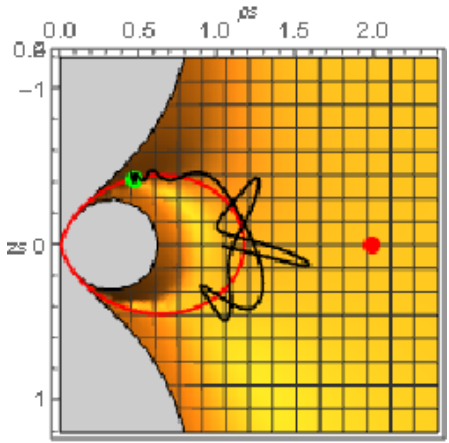

Fig. 4. Downhill orbit for the Størmer problem with initial conditions $\{z_0, \rho_0, p_{z0}, p_{\rho 0}\} = \{0.37,\ 0.4,\ -0.2,\ 0\}$.

The left panel of Fig. 5. shows the maximum width $\Delta\rho$ of a particle's orbit that is caught between two field lines. In the right panel, unexpected jumps in the widths $\Delta\rho$ of orbits for various initial values of $\rho_0$ are predicted.

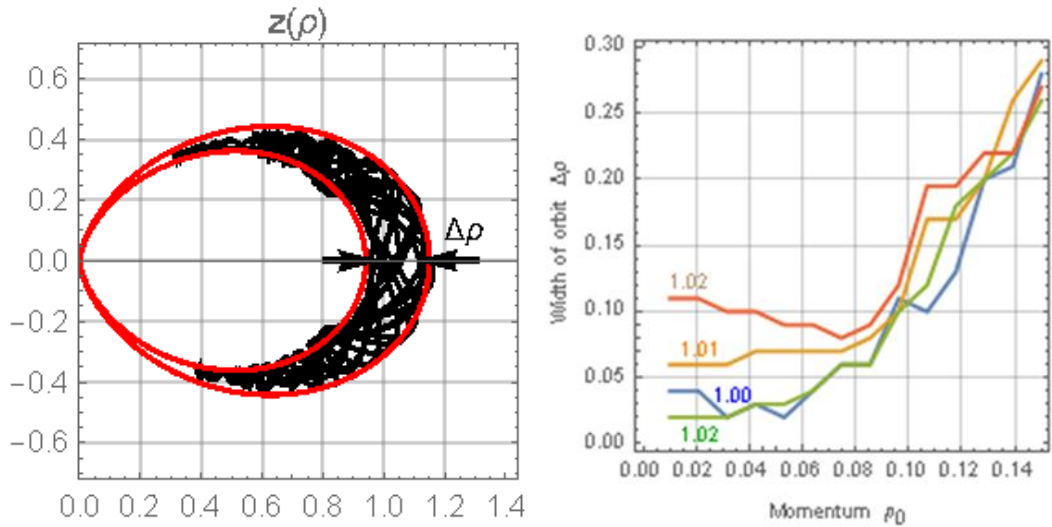

Fig. 5 Left panel: Width $\Delta\rho$ of an orbit with initial condition $\{z_0, \rho_0, p_{z0}, p_{\rho 0}\} = \{0, 1.05, 0.21, 0\}$. Right panel: For low momenta, the width $\Delta\rho$ of an orbit changes strongly with initial condition $\rho_0 = 1.00$, …, to $\rho_0 = 1.02$.

Next I sketch the main features of chaos theory.

## III. CHAOS THEORY AND THE STØRMER PROBLEM

As is well known, a dynamical system is called chaotic when a small change in its initial conditions leads to widely diverging changes in the future trajectory of the system. The rate of separation is characterized by so-called Lyapunov exponents $\Lambda$. For a Hamiltonian system, the sum of the Lyapunov exponents must be zero, and the Lyapunov exponents must be in pairs of opposite signs. For an *n*-dimensional dynamical system, there are up to *n* positive Lyapunov exponents.

In our case of a two-dimensional system with phase space $\{z, \rho, p_z, p_\rho\}$, there are at most two positive Lyapunov exponents. If the maximum Lyapunov exponent is $\Lambda = 0$, then the particle has a quasiperiodic orbit. If there exists one positive Lyapunov exponent, then the particle's orbit is called chaotic, for two positive Lyapunov exponents, it is called hyper-chaotic.

The type of motion of trapped particles depends primarily on its energy. Particles with low energies will have quasiperiodic orbits, those with intermediate energy will move in a chaotic manner with one positive Lyapunov exponent, and high-energy particles will display hyperchaotic motion with two positive Lyapunov exponents.

Hence, chaos theory then gives us five different types of orbits in the Størmer potential *V*:

- For $\Lambda = 0$, some orbits are stable and purely oscillatory (integrable, regular), but their measure is zero.
- All other orbits with $\Lambda = 0$ are in principle unstable, but practically stable (like the earth's orbit about moon and sun, seen as a three-body problem) and are called quasiperiodic.
- For $\Lambda = 1$, the irregular orbits are called chaotic.
- For $\Lambda = 2$, the irregular orbits are called hyperchaotic.
- When energy parameter is $h > 1/32$, the particle escapes over the saddle point to infinity, also called scattering states.

The lines separating the different regions of stability for $\Lambda = 1, 2, 3$ are fractals, while the line separating the $\Lambda = 2$ region from the scattering region is sharp, due to energy conservation.

Before the advent of chaos theory, publications on the Størmer problem were mostly limited to finding regular orbits, simply because the latter were regarded as intractable. The advent of chaos theory first did not change much on this situation, and one had to wait until the turn of the millennium when, to my knowledge, a first Lyapunov coefficient for a special case of the problem was published in Ref. [7]. The reason for this shortcoming may be that calculations of Lyapunov coefficients are very time consuming (expensive) and full of pitfalls [8].

Fig. 6 from the recent Ref. [9] shows the distribution of the number of positive Lyapunov exponents $\Lambda$=0, 1, 2 in the phase space $\{0, \rho, p_z, 0\}$. Red, orange, and yellow correspond to $\Lambda$ = 0, 1, and 2, respectively. The colored convex curves show the energy, indicated by $h$ in the inset.

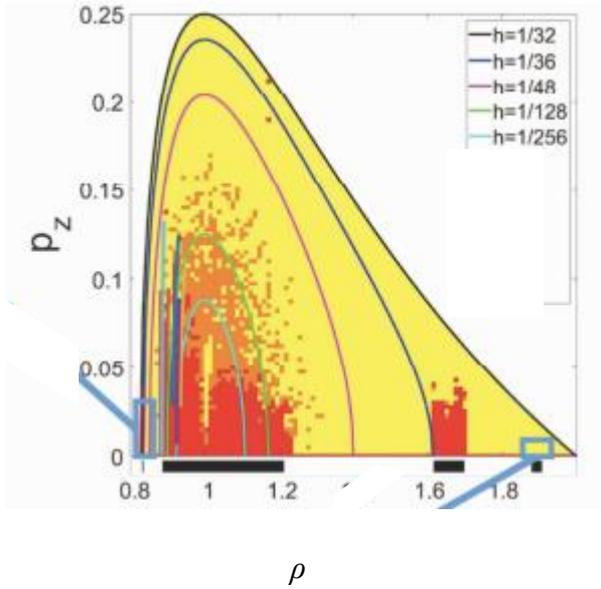

$\rho$

Fig. 6. Distribution of the number $n$ of possible positive Lyapunov exponents in the phase space $\{0, \rho, p_z, 0\}$ of the two-dimensional Størmer problem in the meridian plane. Red for $n = 0$ (quasiperiodic orbits), orange for $n = 1$ (chaotic), and yellow for $n = 2$ (hyperchaotic orbits). The colored convex curves show the energy contours (indicated by $h$ in the inset). From Yuxin Xie and Siming Liu, Chaos **30**, 23108, (2020), Ref. [9].

An open question still is whether the spectral changes due to Størmer, Eq.(6), will lead to changes of beta decay spectra, and with it of the limits on neutrino masses as derived from the KATRIN experiment [10], and on electron-neutrino correlations, as derived from the aSPECT experiment [11]. This will be investigated in a separate paper.